\documentclass[apl,showpacs,preprintnumbers,twocolumn]{revtex4}
\usepackage{amssymb}
\usepackage{color}
\usepackage{graphicx}
\usepackage{graphics}

\begin{document}
\title[GraphiteSchottky]{Graphite based Schottky diodes formed on Si, GaAs and 4H-SiC substrates}
\author{S. Tongay, T. Schumann and A. F. Hebard}
\email[Corresponding author:~]{afh@phys.ufl.edu}
\affiliation{Department of Physics, University of Florida, Gainesville FL 32611}
\author{}

\begin{abstract}
We demonstrate the formation of semimetal graphite/semiconductor Schottky barriers where the semiconductor is either silicon (Si), gallium arsenide (GaAs) or 4H-silicon carbide (4H-SiC). Near room temperature, the forward-bias diode characteristics are well described by thermionic emission, and the extracted barrier heights, which are confirmed by capacitance voltage measurements, roughly follow the Schottky-Mott relation. Since the outermost layer of the graphite electrode is a single graphene sheet, we expect that graphene/semiconductor barriers will manifest similar behavior. 
\end{abstract}

\maketitle
\date{\today}

Metal-semiconductor contacts are ubiquitous in semiconductor technology not only because they are unavoidable, but also because the associated (Schottky) barriers to electronic transport across the metal-semiconductor interface can be tuned by judicious choice of materials and processing techniques\cite{Tung}. The most prominent property of a Schottky barrier is its rectifying characteristic; the barrier acts like a diode with large currents flowing for forward bias and significantly smaller currents flowing for reverse bias\cite{Neamen}. If low resistance and ``ohmic'' (linear) I-V characteristics are desired, then materials and/or processing techniques are chosen to assure that the Schottky barrier height (SBH) $\phi_{B}$ is small compared to temperature (i.e., $\phi_B << k_BT$). Semimetal rather than metal electrodes can also be used. For example, epitaxial ErAs/InAlGaAs diodes fabricated by molecular beam epitaxy have barrier heights that can be tuned over a wide range by adjusting composition and doping\cite{Zimmerman2005}.

Here we report on the use of highly oriented pyrolytic graphite (HOPG) as the semimetal in semimetal/semiconductor Schottky barriers. We demonstrate 
rectifying characteristics on three different $n$-type semiconductors each of which is uniquely suited to specific applications: namely Si, with its robust oxide, to field gated transistors, GaAs, with its direct band gap, to spintronic and optical applications and SiC, with its high thermal conductivity and breakdown strength, to high power/frequency devices. Advantageously the HOPG contact, which can be applied at room temperature, causes minimal disturbance at the semiconductor surface for two reasons: the graphene sheets of the graphite are robustly impervious to diffusion of impurity atoms\cite{Geim2009} and the Van der Waals force of attraction is relatively weak. Since $\phi_{B}$ is related to an interfacial dipole layer associated with bond polarization\cite{Tung}, we infer that barrier properties are determined primarily by the outermost layer of the HOPG contact, i.e., a single layer graphene (SLG) sheet. Accordingly, our results anticipate similar phenomenology using two-dimensional (2D) graphene rather than three-dimensional (3D) graphite. Other examples demonstrating SLG-like properties in graphite include ARPES evidence for the precursor influence of K-point Dirac fermions\cite{Zhou2006} 
and a pronounced temperature-dependent upturn in the in-plane resistivity ($\rho_{ab}$) in the 300~K $< T <$ 900~K temperature range where the next-to-nearest neighbor couplings can be ignored so that the graphite can be described as a stack of graphene bilayers\cite{Dima}.

We have used commercially available \textit{n}-type Si and GaAs with $1\times10^{15}$ cm$^{-3}$ phosphorus (P) and $3\times10^{16}$ cm$^{-3}$ silicon (Si) doping densities respectively. The 4H-SiC wafers are layered, comprising a 5 $\mu$m-thick layer of doped epilayer ($1\times10^{16}$cm$^{-3}$) deposited onto an insulating 4H-SiC substrate. The substrates are thoroughly cleaned to remove any native oxide and/or contaminants. Ohmic contacts are made on the substrates using existing ohmic contact recipes\cite{Hao,Han,Sze}.  

The HOPG contacts are made to the semiconductors using three related techniques:  (1) spring loaded bulk HOPG, (2) Van der Waals adherence of cleaved HOPG flakes or (3) HOPG ``paint''. In the first technique a relatively large ($\sim$1~mm$^2$) piece is gently pressed onto the substrate. In the second cleavage technique, mechanically exfoliated HOPG sheets are landed on the semiconducting substrates. 
Occasionally, relatively large area ($\sim 0.5$~mm$^2$) HOPG flakes flatten out with strong adherence to the substrate due to Van der Waals attraction. In the paint technique, graphite powder/flakes are sonicated in residue-free 2-butoxyethyl acetate and octyl acetate and the painted contacts allowed to air dry. All of these ``soft-landing'' techniques give similar results when the applied currents are normalized with respect to contact area.
 
\begin{figure}[t]
\includegraphics[angle=0,width=0.47\textwidth]{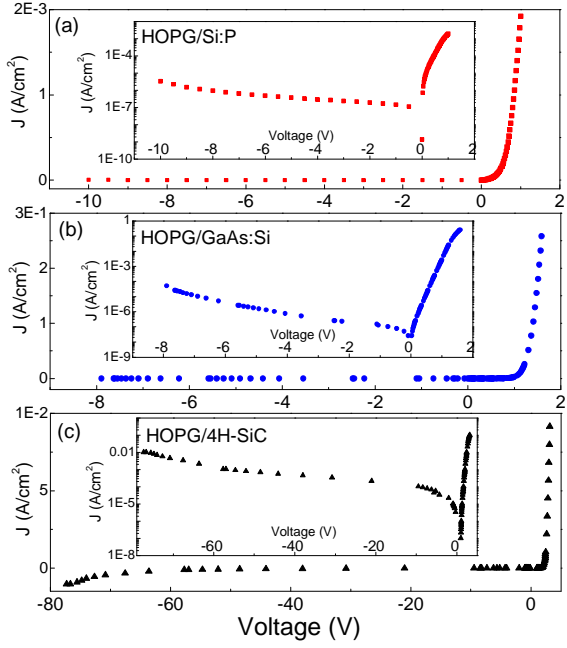}
\caption{\label{IV} Plots of the room temperature current density $J$ with respect to applied bias $V$ on (a) \textit{n}-type Si/graphite (red squares) (b) \textit{n}-type GaAs/graphite (blue circles) and (c) \textit{n} type 4H-SiC/graphite junctions (black triangles). Insets: $J-V$ plots on semilogharithmic axes}
\end{figure}
 
The three panels of Fig.~\ref{IV} show the measured current density vs. voltage ($J$-$V$) room-temperature characteristics of HOPG paint contacts on Si,GaAs and 4H-SiC substrates. These data represent a subset of 27 different samples, all giving similar results independent of the method of application of the HOPG electrode. As seen from Fig.~\ref{IV}, graphite based junctions show good rectification at room temperature. For all of the junctions the rectification is preserved down to 20K.

When electron transport over the barrier height is dominated by thermionic emission, the semilogarithmic  J-V curves usually display a sufficiently linear portion in the forward bias region from which estimates of the zero bias barrier height $\phi_{B0}$ and the ideality constant $\eta$ can be extracted. The extraction is based on the Richardson equation, 
\begin{equation}  \label{richard}
I = I_{s}(T) [\exp ({qV}/{\eta k_B T})-1]   ,
\end{equation}
where $I_{s} = A A^* T^2 \exp ({-q \phi_{B0}}/{k_B T})$
is the saturation current, $q\phi_{B0}$ is the zero bias SBH, $A^{*}$ is the Richardson constant, $T$ is the absolute temperature, and $V$ is the voltage across the ohmic and HOPG contacts. As shown in the panel insets of Fig~\ref{IV}, the HOPG/Si junctions displayed 2-3 decades of linearity in the semilogarithmic J-V curve while the HOPG/GaAs and HOPG/4H-SiC junctions displayed respectively 6 and 4 decades of linearity. The deviations from linearity can be attributed to the existence of more than one transport process, such as space-charge limited emission at low voltages and series resistance effects at higher voltages.

\begin{figure}[b]
\includegraphics[angle=0,width=0.4\textwidth]{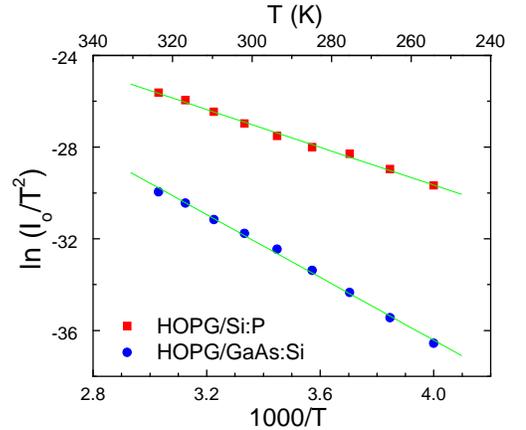}
\caption{\label{activationfig} Richardson activation plots, i.e ($\ln I_{s}/T^{2}$) as a function of $T^{-1}$ from 250~K up to 330~K on (a) HOPG/Si:P (red squares) and (b) HOPG/GaAs:Si junctions (blue circles).}
\end{figure}

Extraction of a reliable value for $\phi_{B0}$ from Eq.~\ref{richard} requires knowledge of the electrically active area $A$. For the HOPG paint contacts, $A$ is not accurately known due to the unknown contact areas of the randomly distributed graphite pieces/flakes on the semiconductor. We remedy this situation by plotting semilogarthmic isothermal $I$-$V$ curves, rather than the $J$-$V$ curves shown in Fig.~\ref{IV}, and then use extrapolation from the linear regions to $V=0$ to determine $I_s(T)$. Analysis is facilitated by writing the equation for $I_s(T)$ in the form,
\begin{equation}
\ln ({I_{s}(T)}/{T^2}) = \ln (AA^*)-({q\phi_{B0}}/{k_{B}T}),   
\end{equation}
where the unknowns $A$ and $\phi_{B0}$ now appear in separate terms.
Typical Richardson activation energy plots of $\ln(I_{s}(T)/T^2)$ versus $T^{-1}$ are shown in Fig.~\ref{activationfig} over the temperature range 250-330~K for HOPG/Si:P (red squares) and HOPG/GaAs:Si (blue circles) junctions. The effective SBHs are calculated from the slopes to be 0.40(1)~eV and 0.50(1)~eV for Si and GaAs respectively (Table~\ref{table}) with ideality factors ($\eta$) spanning from 1.25 to 2.0 for the paint samples shown in Fig.~1. The ideality factors ($\eta$) of the graphite flake samples ($1.12\leq\eta\leq1.50$) are found to be typically smaller than those of the samples prepared by the paint and pressure-contact methods. 

Values of $\eta$ greater than unity are generally attributed to bias dependent SBHs, generation-recombination, thermally assisted tunneling, and image force lowering~\cite{Tung}. These effects can be quantitatively taken into account by finding the flat band zero-electric-field  SBH, $\phi_{BF}$, where surface surface states, if they exist, are depleted of charge and tunneling and image force lowering effects are not present. Theoretical arguments supported by experimental data for $\eta$ in the range $1.05\leq\eta\leq2.2$ validate the relation \cite{Sugerman,Guttler}:
\begin{equation} \label{ideality}
\phi_{BF} = \eta\phi_{B0}-(\eta-1)(k_{B}T/e)\ln(N_C/N_D)
\end{equation}
where $N_D$ and $N_C$ are respectively the doping density and the effective density of states in the conduction band.  Using this expression, the calculated $\phi_{BF}$ values are found to be larger than $\phi_{B0}$ and are closer to the SBH values determined by the C-V measurements (Table~\ref{table}). 

Using the Schottky-Mott relation, $\phi_{BF,B0}=\phi_{m}-\chi$, which relates SBH to the metal work function $\phi_m$ and the semiconductor electron affinity $\chi$, together with the assumption that the Fermi levels of the semiconductors are not pinned, we calculate the HOPG contact work function ($\phi_{HOPG}$) to be in the range 4.40~eV-4.80~eV (Table~\ref{table}). Although the HOPG/4H-SiC junctions do not reveal comparable linearity in the activation energy plots, we can still estimate SBHs using Eq.~\ref{richard}-~\ref{ideality} and with the contact area and theoretical value of the Richardson constant by fitting the $J-V$ curves in panel (c) of Fig.~\ref{IV}(Table~\ref{table}). Our values of $\phi_{HOPG}$ determined separately on Si, GaAs and 4H-SiC are in good agreement with the theoretically and experimentally determined values (ranging from 4.4~eV to 4.8~eV) reported in the literature\cite{Suzukia,Briddon,Taft}.

 \begin{table} [t]
 \caption{Extracted SBHs, doping densities, and corresponding graphite work function values on various graphite/semiconductor junctions}
 \centering
 \begin{tabular} {p{2.2cm} p{0.8cm} p{0.8cm} p{0.8cm} p{1.0cm} p{1.0cm} p{0.8cm}}
 \hline\hline
  & $\phi_{Bo}$ & $\phi_{BF}$ &$\phi_{C-V}$ & $N_{D}^{C-V}$ & $N_{D}^{Hall}$ & $\phi_{HOPG}$ \\ [1.0 ex] 
  junction type & [eV] & [eV] & [eV] & [cm$^{-3}$] & [cm$^{-3}$] & [eV] \\ [1.0 ex]
 \hline
 \footnotesize HOPG/\textit{n}Si & 0.40 & 0.60 & 0.70 & 1.2E15 & 1.0E15 & 4.60 \\
 \scriptsize HOPG/\textit{n}GaAs & 0.60 & 0.78 & 0.76 & 3.6E16 & 3.0E16 & 4.78 \\
 \scriptsize HOPG/\textit{n}4H-SiC & 1.15 & 1.60 & 1.84 & 1.2E16 & 1.0E16 & 4.80 \\
 \hline
 \end{tabular}
 \label{table}
 \end{table}

\begin{figure}[b]
\vspace{-10pt}
\includegraphics[angle=0,width=0.45\textwidth]{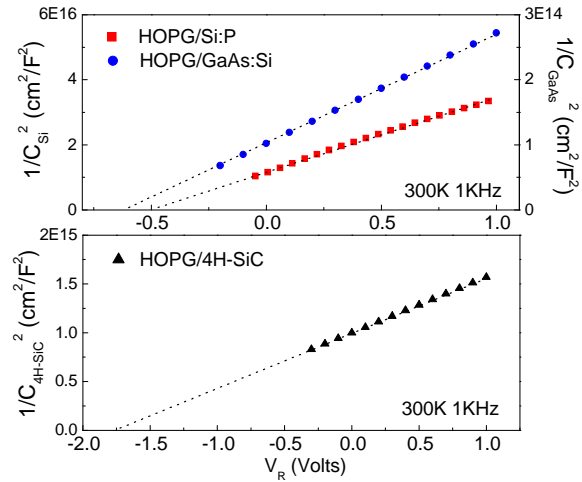}
\caption{\label{capacitance}
Inverse square of capacitance per unit area measured at 1~kHz as a function of reverse bias at room temperature:(top panel) HOPG/Si:P (red squares, right hand axis) and HOPG/GaAs:Si (blue circles, left hand axis); (bottom panel) HOPG/4H-SiC (black triangles).}
\end{figure}

As shown in Fig.~\ref{capacitance}, we have also used capacitance(1~kHz)-voltage ($C$-$V$) measurements plotted in the form $1/C^2$ vs. $V_R$, where $V_R$ is the reverse bias voltage, to characterize our junctions at room temperature. The observed linearity suggests that gap states are absent and that the surface density of states is small\cite{Fonash}. Linear extrapolation (dotted lines) to the intercept with the absiccsa identifies the built-in potential, $V_{bi}$, which is related to SBH via the expression, $\phi_{C-V}=V_{bi}+(E_{c}-E_{F})$, where $E_c$ is the conduction band edge and $E_F$ the Fermi energy. In like manner, the dopant densities of each semiconductor can be calculated from the slopes. The values for $\phi_{C-V}$ and $N_D$ extracted from the linear dependences shown in Fig.~\ref{capacitance} are listed in Table~\ref{table}. Although the values for $N_D$ are in good agreement with Hall data, the values for $\phi_{C-V}$ are observed to be slightly higher than the values extracted from $I$-$V$ measurements. This trend might be attributed to the presence of a very thin oxide layer at the metal semiconductor interface causing $V_{bi}$, and hence $\phi_{C-V}$, to be overestimated by $C$-$V$ measurements ~\cite{Tung}.

In conclusion, we have demonstrated the formation of 
Schottky contacts using a ``soft-landing'' HOPG contact on \textit{n}-type Si, GaAs and 4H-SiC semiconducting substrates. Fabrication can be as easy as allowing a dab of HOPG paint to air dry on any one of the investigated semiconductors. 
The extracted values of SBH from \textit{I-V} and \textit{C-V} measurements roughly obey the Schottky-Mott relation with inferred graphite work functions agreeing well with literature values. Our results not only provide unexpected insights into the nature of the graphite/semiconductor interface but also anticipate applications where single-layer graphene is directly contacted to a semiconductor substrate rather than isolated by an insulating oxide\cite{Novoselov666} or grown directly on undoped insulating semiconductors \cite{deHeer}.

We thank B. Gila and S. Pearton for useful discussions. This research was supported by the NSF under grant numbers DMR-0704240 (AFH) and DMR-0851707 (TS, REU student).

\end{document}